\documentstyle[prb,aps,epsf,floats,goodfloat]{revtex}

\begin{document}
\draft

\twocolumn[\hsize\textwidth\columnwidth\hsize\csname@twocolumnfalse%
\endcsname

\title{
Critical Behavior of the Two-Dimensional Random Quantum Ising Ferromagnet
}

\author{C. Pich and A. P. Young}
\address{Department of Physics, University of California, Santa Cruz, 
CA 95064}

\date{\today}

\maketitle

\begin{abstract}
We study the quantum phase transition in the
two-dimensional random Ising model in a transverse field by Monte Carlo
simulations. We find results similar to those known analytically in
one-dimension:
the dynamical exponent is infinite and, at the critical point,
the typical correlation function decays with
a stretched exponential dependence on distance. Away from the critical point,
there may be different exponents for the divergence of the
average and typical correlation lengths, again as in one-dimension, but the
evidence for this is less strong.
\end{abstract}

\vskip 0.3 truein
]

\section{Introduction}

Though {\em classical} phase transitions occurring at finite
temperature are very well understood, our knowledge of {\em quantum}
transitions at $T=0$ is relatively poor, at least for systems with quenched
disorder. There is, however, considerable interest in these systems since
they (i)
exhibit new universality classes, and (ii) display
``Griffiths-McCoy''\cite{griffiths,mccoy}
singularities even away from the critical point, due to rare regions 
with stronger than average interactions.

Just as the simplest model with a
classical phase transition is the regular Ising model,
the simplest random model with a quantum transition is
arguably the Ising model in a transverse field whose Hamiltonian is given by
\begin{equation}
{\cal H} = -\sum_{\langle i,j\rangle} J_{ij} \sigma^z_i \sigma^z_{j} -
\sum_i h_i \sigma^x_i \ .
\label{ham}
\end{equation}
Here the $\{\sigma^\alpha_i\}$ are Pauli spin matrices, and the
nearest neighbor
interactions $J_{ij}$ and transverse fields $h_i$ are both independent
random variables. 
Naturally this model has been quite extensively studied and many
{\em analytical} results are
available\cite{dsf,sm,mw,mccoy} for the case of the one-dimensional chain.
Among the surprising predictions\cite{dsf} for the critical behavior
in 1-d are:

\begin{enumerate}

\item
\label{point1}
The dynamic critical exponent, $z$, is infinite.
Instead of a
characteristic time scale $\xi_\tau$ varying as a power of
a characteristic length scale $\xi$ according to $\xi_\tau \sim \xi^z$,
one has instead an exponential relation
\begin{equation}
\xi_\tau \sim \exp ({\rm const.}\ \xi^\mu),
\end{equation}
where $\mu = 1/2$. This is called {\em activated} dynamical scaling.

\item
\label{point2}
Distributions of the equal-time
$\sigma^z$-$\sigma^z$ correlations, are very broad. As a result
{\em average} and {\em
typical}\/\cite{typical} correlations behave rather differently,
since the average is dominated by a few rare (and hence {\em
atypical}\/) points. At the critical point, for example, the average correlation
function falls off with a power of the distance $r$ as
$
C_{\rm av}(r) \sim r^{-\lambda} ,
$
where $\lambda = 2-\phi \simeq 0.38$ with $\phi$ the golden mean, whereas the
typical value falls off much faster, as a stretched exponential
\begin{equation}
C_{\rm typ}(r) \sim \exp (- {\rm const. }\ r^\sigma ) ,
\end{equation}
with $\sigma = 1/2$.

\item
\label{point3}
Away from the critical point the average and typical correlation {\em lengths}
both
diverge but with {\em different} exponents, i.e.
$
\xi_{\rm av} \sim \delta^{\nu_{\rm av}}; \ \xi_{\rm typ} \sim \delta^{\nu_{\rm typ}},
$
where $\delta$ is the deviation from criticality and
\begin{equation}
\nu_{\rm av} = 2; \quad \quad \nu_{\rm typ} = 1.
\end{equation}

\end{enumerate}

Unfortunately, the analytical approach is only valid in 1-d, and very little
is known about the critical behavior in higher dimensions\cite{ss}.
An important question is whether these striking analytical results 
are a special feature
of 1-d, or whether they are valid more generally. Here,
we make a first
step in investigating this question by 
performing large-scale Monte Carlo simulations of 
the random transverse field Ising {\em ferromagnet} in two dimensions. Because
the ferromagnet has no frustration we are able to use highly efficient cluster
algorithms\cite{sw,wolff} which considerably reduce critical slowing down.
Consequently, we are able to study larger sizes and get much
better statistics than in earlier work\cite{ry,gbh} on quantum {\em spin
glasses}. Furthermore, cluster algorithms have an improved
estimator\cite{evertz} for correlation functions, which greatly reduces the
statistical errors when the mean is small, so we are able to
probe the distribution of correlation functions out to relatively large
distances.

Our main conclusion is that the behavior of the 2-d system at the
critical point is {\em very similar}
to that of the 1-d case, see points \ref{point1} and \ref{point2}
above. There are some difficulties in interpreting our data away from the
critical point but, nonetheless, here too there is some evidence for behavior
similar to that in 1-d, see point \ref{point3} above. Further details and
additional results will be presented elsewhere\cite{py}.

In related work, Rieger
and Kawashima\cite{rk} have studied the same model but with {\em continuous}
imaginary time, concentrating on
Griffiths-McCoy singularities in the paramagnetic phase, which
are characterized by a
continuously varying dynamical exponent, $z(\delta)$. Kawashima and Rieger find
that $z(\delta)$ tends to infinity as the critical point is approached
($\delta\to 0$), again as in 1-d\cite{dsf}.

As is standard\cite{kogut}, we represent the $d$-dimensional
quantum Hamiltonian in
Eq.~(\ref{ham}) by an effective classical ``Hamiltonian '' in
($d$+1)-dimensions, where the extra dimension, imaginary time, is of size
$\beta \equiv 1/T$ and is divided up into $L_\tau$ intervals each of width
$\Delta \tau = \beta/L_\tau$. The effective classical Hamiltonian is given by
\begin{equation}
\beta_{cl} {\cal H}_{cl} = -\sum_{\langle i, j\rangle, \tau} K_{ij}
S_i(\tau) S_j(\tau) - \sum_{i, \tau} \widetilde{K}_i S_i(\tau) S_i(\tau^\prime)
\end{equation}
where, $\tau^\prime = \tau + \Delta \tau$, $S_i(\tau)=\pm 1$,
$
K_{ij} = \Delta \tau J_{ij},$ $\exp(-2\widetilde{K}_i) =
\tanh(\Delta \tau h_i) ,
$
and $\beta_{cl} \equiv 1/T_{cl}$ where $T_{cl}$
is an effective ``classical'' temperature (not
equal to the real temperature which is the inverse of the size in the time
direction). Note that the disorder is quenched and so the interactions are
independent of $\tau$. Correlations of the classical Ising spins, $S_i(\tau)$,
correspond to $\sigma^z$--$\sigma^z$ correlations in the original quantum
model. These are the correlations that we are interested in here.

In order to capture the random quantum critical behavior in the
intermediate size
systems that we can simulate,
we wish the disorder to be effectively quite strong. In particular, we
would like some of the fields to be {\em much} stronger than the bonds in their
vicinity and vice-versa. This is captured by having distributions
for both the fields and interactions with a finite weight at the origin.
This is satisfied by
a constant distribution for the $K_{ij}$ and a distribution for
the $\widetilde{K}_i$ which falls off exponentially at {\em large} values.
Furthermore, the critical behavior does not depend on short
distance physics and is hence independent of $\Delta \tau$, so
for convenience we will take $\Delta \tau = 1$. A Hamiltonian
which encapsulates
these features in a way convenient for simulations is
\begin{equation}
{\cal H}_{cl} = -\sum_{\langle i, j\rangle, \tau} J_{ij}
S_i(\tau) S_j(\tau) - \sum_{i, \tau} \widetilde{J}_i S_i(\tau) S_i(\tau + 1) 
,
\end{equation}
where $\tau$ runs over integer values, $1\le \tau \le L_\tau$, and the
distributions of the interactions are given by
\begin{eqnarray}
\pi(J) & = &
\left\{
\begin{array}{ll}
1 & \mbox{for $ 0 < J < 1$,} \\
0  & \mbox{otherwise},
\end{array}
\right.
\nonumber \\
\rho(\widetilde{J}) & = &
\left\{
\begin{array}{ll}
2 \exp(-2 \widetilde{J})  & \mbox{for $  \widetilde{J} \ge 0$,} \\
0  & \mbox{for $  \widetilde{J} < 0$.}
\end{array}
\right.
\label{dist}
\end{eqnarray}
Disorder averages are denoted by $[\cdots]_{\rm av}$
and Monte Carlo averages for a single sample are denoted by $\langle \cdots
\rangle$.
One controls the strength of the fluctuations, and hence tunes through the
transition, by varying the classical temperature, $T_{cl}$.

To assess the accuracy of the method, we not only performed simulations
in 2-d,
but also
carried out simulations for the 1-d model for a similar range of sizes
and compared the results
with the analytic predictions discussed above. 
The size of the system in the space direction denoted by
$L$, so there are $N=L^d
L_\tau$ sites in the sample. We studied sizes up to $L=48$ and $L_\tau=2048$
using the Wolff\cite{wolff} cluster algorithm. We found that no more than 100
sweeps were required for equilibration, even for the largest lattices. At least
1000 samples were averaged over.

\begin{figure}
\epsfxsize=\columnwidth\epsfbox{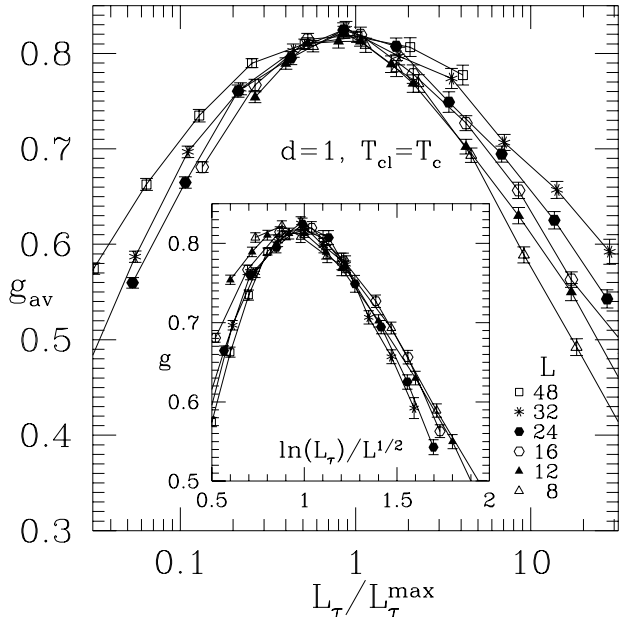}
\caption{
A plot of the data for the average Binder ratio, $g_{\rm av}$, at the critical
temperature $T_{cl} = T_c = 0.98$ for the one-dimensional model. In the main
figure the horizontal axis is
$L_\tau/L_\tau^{\rm max}$ where $L_\tau^{\rm max}$ is
the value of $L_\tau$ at the peak. Note that the curves do not scale but
rather get broader for larger
sizes, indicating activated scaling, $z=\infty$, (which is known from the
analytical work\protect\cite{dsf}). The inset shows a plot appropriate for
activated scaling in which the horizontal axis is $\ln L_\tau / \ln
L_\tau^{\rm max}$, with $\ln L_\tau^{\rm max} \sim L^{1/2}$,
which is also known from the
analytical work. Scaling works quite well though there are still some
small deviations for this range of sizes, even in the large $L_\tau$ limit.
}
\label{g-1d}
\end{figure}

\begin{figure}
\epsfxsize=\columnwidth\epsfbox{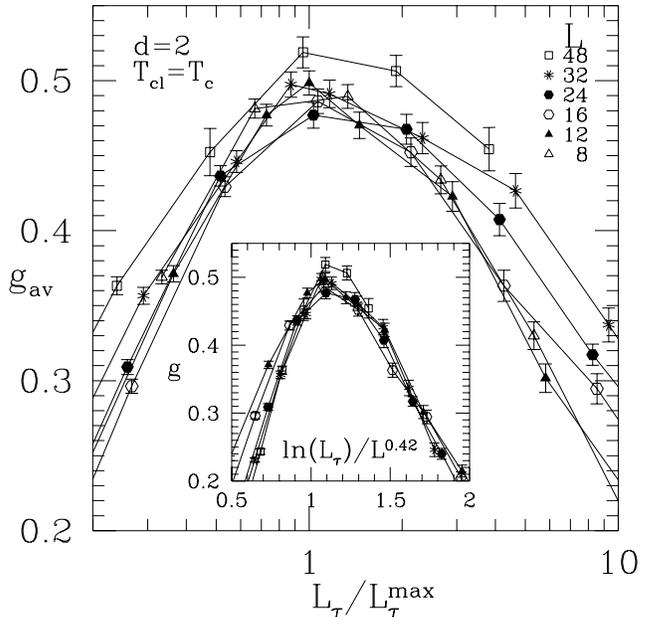}
\caption{
As in Fig.~\protect\ref{g-1d} but for two dimensions, with $T_{cl} = T_c
= 2.45$. The conclusion
is similar, i.e. the broadening of the curves for large $L$ in the
main plot indicates activated dynamical scaling, i.e. $z=\infty$. The
data for $L=48$ is slightly high which may indicate that the true value of
$T_c$ is a little higher. In the inset, the data for $L \ge 12$ is seen to
scale quite
well with the same form $\ln L_\tau  / L^{\mu}$ known to be exact in 1-d, but
the value of $\mu=0.42$ (shown)
works better than the 1-d value of $\mu=1/2$. 
}
\label{g-2d}
\end{figure}

The first task is to locate the critical point, for which we use a method
suggested by D.~Huse (private communication)\cite{ry,gbh}.
One computes the Binder ratio
\begin{equation}
g_{\rm av} = {1 \over 2} \left[ 3 - {\langle M^4\rangle \over \langle M^2\rangle^2}
\right]_{\rm av} , 
\end{equation}
where $M = \sum_{i,\tau} S_i(\tau)$,
which (assuming, for now, that $z$ is finite) has the finite-size scaling form
\begin{equation}
g_{\rm av} = \widetilde{g}\left( \delta L^{1/\nu_{\rm av}}, L_\tau/L^z \right) .
\end{equation}
For fixed $L$, $g_{\rm av}$
has a peak as a function of $L_\tau$. At the critical
point, $T_c$\cite{tc}, the peak height is independent of $L$ and the values of
$L_\tau$ at the maximum, $L_\tau^{\rm max}$, vary as $L^z$. Furthermore,
a plot of $g_{\rm av}$ against $L_\tau/L_\tau^{\rm max}$ at the critical point,
which has the advantage of not needing a value for $z$,
should collapse the data. We see in Figs.~\ref{g-1d} and \ref{g-2d}
that this does {\em not}
happen either in 1-d or 2-d. Rather the curves clearly
become broader for larger sizes.
This is easy to understand since we know that for 1-d $z$ is infinite and 
it is the
{\em log} of the characteristic time which
scales with a power of the length scale.
This
suggests that the scaling variable should be $\ln L_\tau / \ln L_\tau^{\rm max}$
with $\ln L_\tau^{\rm max} \sim L^\mu$, say. The inset to Fig.~\ref{g-1d}
shows that this works
moderately well for 1-d (though not perfectly for this range of sizes) with the
expected value $\mu=1/2$. In 2-d, the
data collapse for sizes $L \ge 12$,
shown in the inset to
Fig.~\ref{g-2d}, is quite good for $\mu=0.42$, and not quite so good with
the 1-d value, $\mu=1/2$, though we would not claim that $\mu=1/2$ is ruled
out.
The close
similarity of the data for 1-d and 2-d, particularly the broadening of
the data in the
main part of the figures, suggests that $z$ is infinite also in 2-d.

\begin{figure}
\epsfxsize=\columnwidth\epsfbox{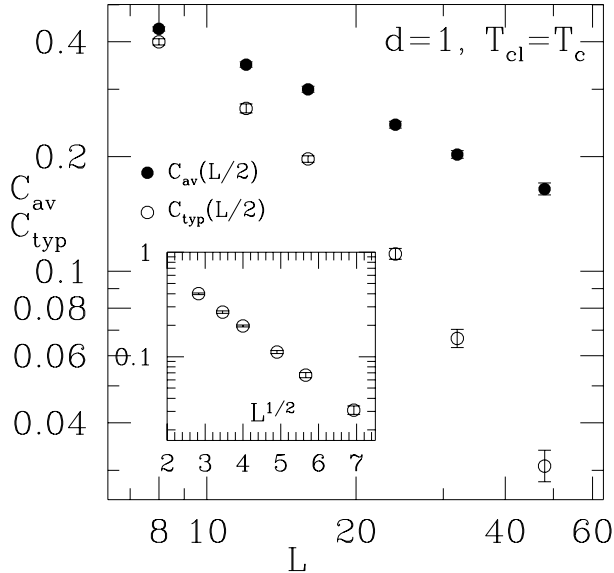}
\caption{
The main figure shows the average and typical\protect\cite{typical} correlations
between spins $L/2$ apart at the critical point in
one dimension. As expected\protect\cite{dsf},
the average decays with a power law (though the fit
to the data gives a slope, $-\lambda$, with $\lambda=0.50$
compared with the analytical value
of $0.38$),
while the typical value
decays faster than a power, as shown by the downward
curvature in the plot. The inset shows that the typical correlation function
decays as the stretched exponential, $C_{\rm typ}(L/2) \sim \exp (-{\rm const.}\
L^{1/2})$, as expected\protect\cite{dsf}.
}
\label{cf-1d}
\end{figure}
\begin{figure}
\epsfxsize=\columnwidth\epsfbox{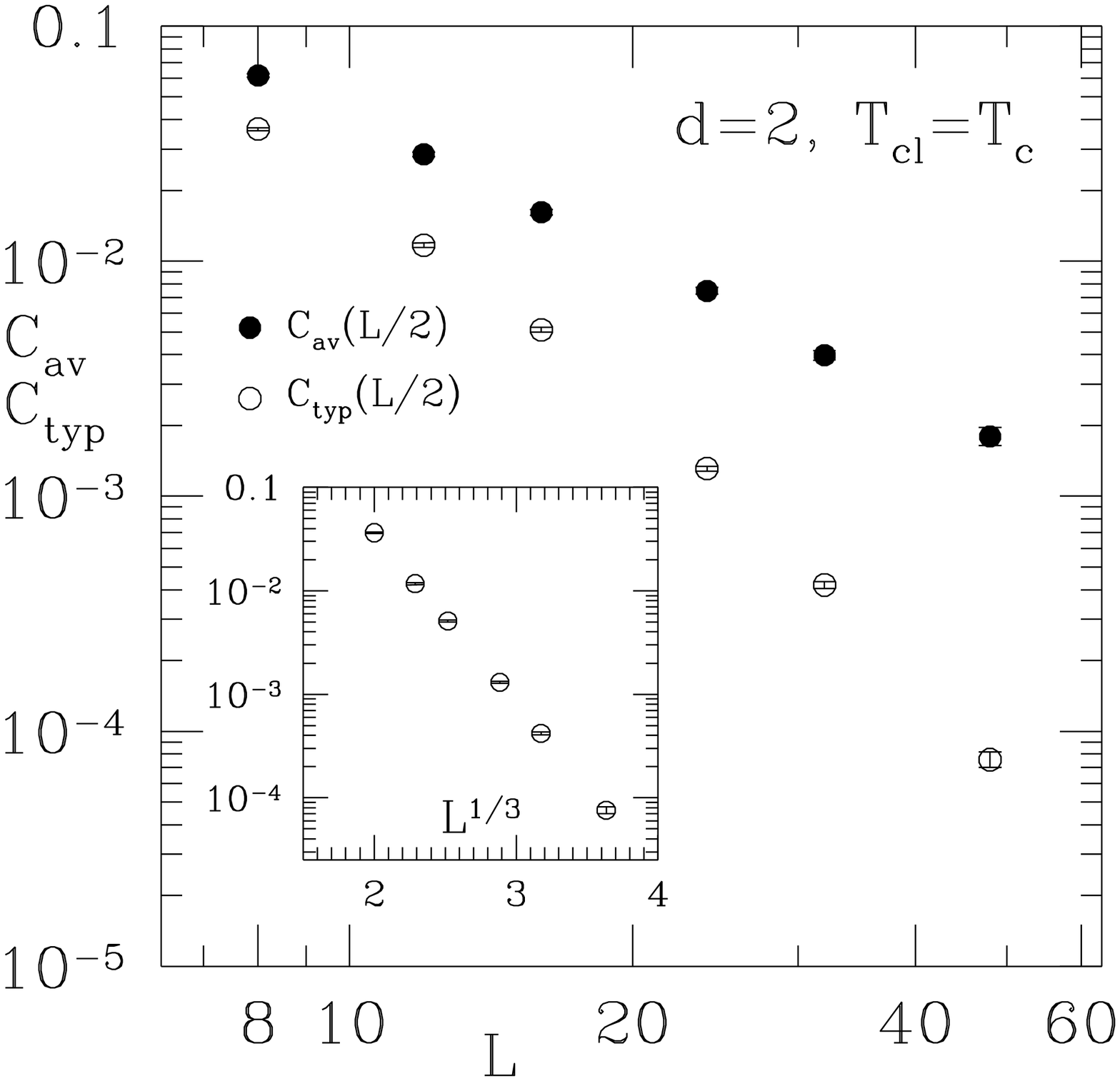}
\caption{
A similar plot to Fig.~\protect\ref{cf-1d} but for two dimensions. We emphasize
just how similar these figures are. The most noteworthy feature is that in two
dimensions, as in d=1, the typical correlation function falls off {\em faster}
than a power law. The average falls off with a power law and a fit
gives a slope of $-\lambda$ with $\lambda=1.50$.
The inset shows that the 2-d 
data is consistent with a stretched
exponential form, $\exp (-{\rm const.}\ L^\sigma)$, but with $\sigma=1/3$
rather than the value of 1/2 found in 1-d.
}
\label{cf-2d}
\end{figure}

Next we consider the behavior of the equal time correlations at the critical
point using
the improved estimator\cite{evertz}. Figs. \ref{cf-1d} and
\ref{cf-2d} show data for the
average and typical\cite{typical} correlations for spins separated by 
$r=L/2$ in 1-d and $\vec{r} = (L/2, 0)$ in 2-d. For each value of $L$, we took
$L_\tau$ such that $g_{\rm av}$ is close to the peak\cite{L-Ltau},
see Figs.~\ref{g-1d} and ~\ref{g-2d}.
According to finite size
scaling, the dependence on $L$ for a finite system should be the same as the
dependence on $r$ in a bulk system. Fig.~\ref{cf-1d}
shows that in 1-d, the
average correlation falls off with a power law while the typical
value decays with an
$\exp(-{\rm const.}\ L^{1/2})$ variation, as expected\cite{dsf}.
Note, though, that the exponent for the average, $-\lambda$,
obtained by fitting the data,
gives $\lambda=0.50$, somewhat
different from the expected\cite{dsf}
value of $0.38$. We will discuss this
discrepancy further below. The data in Fig.~\ref{cf-2d}
shows that the behavior in 2-d is
quite similar: the average falls off with a power law, with $\lambda$ about
$1.5$, while the typical value falls off faster than a power
law, (because of the downward {\em curvature}) consistent with a
stretched exponential behavior of the form $\exp(-{\rm const.}\ L^\sigma)$,
with $\sigma \simeq 1/3$. The statistical errors (shown) are generally smaller
than the size of the points, so the downward curvature 
is statistically significant.
The value of $\sigma$ seems to be different from the 1-d result of
$1/2$ (the data is curved on the appropriate plot) 
but we are not able to determine it with great precision.

Finally we briefly describe our data away from the critical point, keeping
keep the same values of $L$ and $L_\tau$ as at the critical point.
In 1-d, $g_{\rm av}$ scales well with the
expected value $\nu_{\rm av}=2$
and the corresponding plot in 2-d also works well,
this time with $\nu_{\rm av}\simeq 1.5$.
Plots of the typical correlation function (divided by the value at criticality)
give an exponent close to the expected value of $\nu_{\rm typ} = 1$ in 1-d and
around 0.7 in 2-d, i.e. in each case {\em much smaller}
than the exponent for the
average correlation length obtained from $g_{\rm av}$.
However, attempts to determine
$\nu_{\rm av}$ {\em directly}
by scaling data for the average correlation function
away from criticality
were less successful.
In 1-d, the fit, which works quite well, gives $\nu_{\rm av}
\simeq 1.2$, much less than the expected value\cite{dsf} of 2,
while in 2-d we found
$\nu_{\rm av} \simeq 0.75$, not significantly different from $\nu_{\rm typ}$.

This discrepancy for results from average correlation function
may be due to
difficulties in taking enough samples to probe the rare regions which dominate
the average. This may also be why the exponent for the decay of the average
correlation at criticality in 1-d was not found accurately.
Another concern with the data away from the critical point is that we do not
understand
the way that the critical
singularities go over to Griffiths-McCoy singularities,
which have a continuously variable\cite{dsf,rk,yr} $z(\delta)$.
Perhaps
the latter give big corrections to scaling, thus possibly making the size of
the critical region quite small.

To conclude, we have found a strong similarity between the critical behavior of
the random transverse field Ising model in one and two dimensions. 
The evidence that $z=\infty$ and that
the typical correlations at the
critical point decay with a stretched exponential function of distance are
quite strong, but the
evidence for different exponents for the average and typical
correlations in 2-d is weaker because of discrepancies involving the
data for the average correlation function.

After this work was completed, we heard that
S.-C. Mau, O. Motrunich and D. A. Huse (private communication)
have implemented numerically for $d > 1$
the renormalization group approach used in Ref. \onlinecite{dsf}, finding
a flow to the infinite disorder
critical fixed point, just as in $d=1$.

\acknowledgments
We would like to thank D.~S.~Fisher, D.~A.~Huse, H.~Rieger, N.~Kawashima, 
and R.~N.~Bhatt for helpful discussions, and O.~Narayan for a critical
reading of the manuscript. 
This work was supported by the National Science Foundation under
grant DMR 9713977 and the Deutsche Forschungsgemeinschaft (DFG) under contract
Pi 337/1-2.

\end{document}